\documentclass[showpacs,twocolumn,prl]{revtex4}
\usepackage{graphicx,amssymb}
\newcommand{\ket}[1]{\left|#1\right\rangle}
\newcommand{\bra}[1]{\left\langle#1\right|}
\newcommand{\op}[2]{\ket{#1}\bra{#2}}

\newcommand{\proj}[1]{\op{#1}{#1}}
\newcommand{\com}[2]{\left[#1,#2\right]}
\newcommand{\brac}[1]{\left\{#1\right\}}

\newcommand{\brar}[1]{\left(#1\right)}

\newcommand{\dampingnobr}[3]{2#1#2#3-#3#1#2-#2#3#1} 
\newcommand{\damping}[3]{\brar{\dampingnobr{#1}{#2}{#3}}}
\newcommand{\modulus}[1]{\left|#1\right|}
\newcommand{\modsquared}[1]{{\modulus{#1}^2}}

\newcommand{\commentout}[1]{}
\newcommand{\mat}{\modulus{a}^2}
\newcommand{\maf}{\modulus{a}^4}
\newcommand{\mas}{\modulus{a}^6}
\newcommand{\mbt}{\modulus{b}^2}
\newcommand{\mbf}{\modulus{b}^4}
\newcommand{\mbs}{\modulus{b}^6}
\newcommand{\ep}{\epsilon}

\begin{document}

\title{Unconditional preparation of entanglement between atoms in
cascaded optical cavities}
\author{Stephen Clark}
\author{Amy Peng}
\author{Mile Gu}
\author{Scott Parkins}
\affiliation{Department of Physics, University of Auckland,
Private Bag 92019, Auckland, New Zealand.}
\date{\today}

\begin{abstract}
We propose a scheme to unconditionally entangle the internal
states of atoms trapped in separate high finesse optical cavities.
The scheme uses the technique of quantum reservoir engineering
in a cascaded cavity QED setting, and for ideal (lossless)
coupling between the cavities generates an entangled {\em pure}
state. Highly entangled states are also shown to be possible
for realizable cavity QED parameters and with nonideal coupling.

\end{abstract}
\maketitle

Cold trapped atoms and quantum light fields are promising
candidates for the realization of quantum computing and
quantum communication protocols \cite{Monroe02,Mabuchi02},
with long-lived atomic states (electronic or motional)
constituting quantum registers, upon which (local) quantum
logic operations can be performed, and light fields
providing a means of distributing quantum information and
entanglement between different nodes in a network of
registers \cite{Cirac97}. The workability of such atom-light
networks will depend heavily on the extent to which
propagating light fields can reliably transfer quantum states
and/or establish quantum entanglement between
atoms at different nodes of the network.

In the context of entanglement preparation between atoms at
separate nodes, a variety of schemes have been
proposed recently. Based on their operating
principles, these schemes can be grouped loosely as follows:
(i) ``Local'' entanglement, prepared by some means between
atoms at one node, is transferred, via carefully-controlled
quantum state-transferring light pulses, from a subset of the
entangled atoms to atoms at a distant node
\cite{Cirac97,Parkins01}.
(ii) Quantum-correlated light fields, produced, e.g., by
nondegenerate parametric downconversion, interact with
separate atoms in such a way as to transfer some of
their properties to, and thereby entangle, the atoms
\cite{Polzik99,Kuzmich00,Parkins00,Molmer00,Lloyd01,Lukin00}.
(iii) Measurements (e.g., single-photon detections or homodyne
detection over some interval) are made on superpositions of
light fields emanating from separate atomic samples, or on a
probe light field that has interacted in a prescribed way with
different samples. Indistinguishability in the measurement
conditionally projects the atomic systems into an entangled
state
\cite{Cabrillo99,Bose99,Feng03,Duan03,Simon03,Plenio03,Gilchrist02,%
Duan00,Duan01,Duan02}.

Here we propose a scheme for preparing distributed
atomic entanglement that is quite distinct from those listed
above. While it employs cascaded cavity QED systems (as, e.g.,
in \cite{Cirac97,Parkins01}), it does not require initial local
entanglement between atoms or tailored optical pulses, nor does
it involve separate nonclassical light sources or projective
measurements. The entangled atomic state is prepared
{\em unconditionally} and under {\em steady state} conditions.
Furthermore, the degree of entanglement (and also the mixedness)
of the state is adjustable through variation of
certain tunable system parameters (i.e., Raman coupling
strengths), and, for ideal transmission of light between
cavities, a {\em pure} entangled state can be prepared.

\begin{figure}
\begin{center}
\includegraphics[scale=0.55]{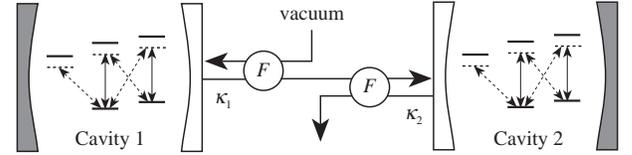}
\caption{Cascaded cavities, each containing a trapped
atom. A unidirectional coupling between the
cavities is achieved using Faraday isolators ({\em F}).
The (one-sided) cavities have field decay rates~$\kappa_1$
and~$\kappa_2$.}
\label{fig:cavities}
\end{center}
\end{figure}

Our scheme employs quantum reservoir engineering
\cite{Lutkenhaus98} in a cavity QED setting \cite{Clark03}.
Two high-finesse optical cavities, each containing
one tightly confined atom, are arranged in a cascaded
configuration with a unidirectional coupling from cavity 1
to cavity 2 (Fig.~\ref{fig:cavities}).
Both cavities are taken to have the same resonant frequency
$\omega_{\rm cav}$ and their individual field decay rates
are $\kappa_1$ and $\kappa_2$. Inefficiencies and losses in
the coupling between the two cavities are modelled by a real
parameter $\epsilon$, where $0\le\epsilon\le1$ and ideal
coupling corresponds to $\epsilon=1$.

\begin{figure}
\begin{center}
\includegraphics[scale=0.75]{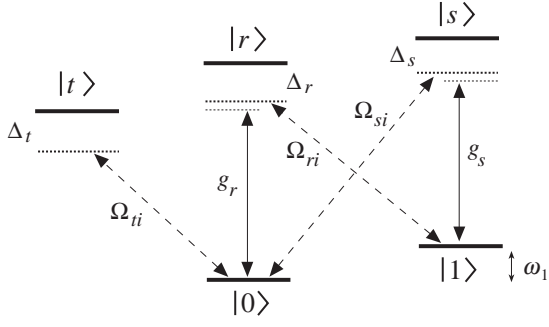}
\caption{Level scheme for each atom. The excited states
have energies $\hbar\omega_j$ ($j=r,s,t$). Such an atomic
configuration could be realized, e.g., with alkali atoms,
where $|0\rangle$ and $|1\rangle$ are different ground-state
sublevels. Note also that $|r\rangle$ and $|s\rangle$
can be the same level, provided the two Raman channels remain
distinct (which would require $\omega_1\neq 0$).
Apart from $\Omega_{ri}$, $\Omega_{si}$ and $\Omega_{ti}$,
we assume, for simplicity, that all other parameters are the
same for each atom.}
\label{fig:levels}
\end{center}
\end{figure}
Each atom has two stable ground states, $\ket{0}$ and $\ket{1}$
(the qubit states).
The cavity field and two auxiliary laser fields drive two
separate Raman transitions between these states
(Fig.~\ref{fig:levels}). In particular,
transitions $\ket{1}\leftrightarrow\ket{r}$ and
$\ket{0}\leftrightarrow\ket{s}$ are driven by detuned laser
fields with (complex) Rabi frequencies $\Omega_r$ and $\Omega_s$,
while the transitions $\ket{0}\leftrightarrow\ket{r}$ and
$\ket{1}\leftrightarrow\ket{s}$ are strongly coupled to the
cavity mode, with coupling strengths $g_r$ and $g_s$.
Detunings of the fields from the
excited states $\ket{r}$ and $\ket{s}$ are $\Delta_r$ and
$\Delta_s$. A fifth state $\ket{t}$ is virtually excited from
$\ket{0}$ by another strongly detuned laser field, adding an
additional ac-Stark shift to the state $\ket{0}$.

The master equation for the total system density operator
$\rho_{\rm T}$ is (taking $\hbar =1$)
\begin{equation}
  \dot{\rho}_{\rm T}=-i\com{H}{\rho_{\rm T}}
  +\mathcal{L}_{\rm cav}\rho_{\rm T}
  +\mathcal{L}_{\rm spon}\rho_{\rm T} ,
\end{equation}
where $H=H_{\rm cav}+H_{\rm at}+H_{\rm int}$, with
\begin{eqnarray}
H_{\rm cav}&=& \sum_{i=1,2} \omega_{\rm cav} a_i^\dag a_i \, ,
\\
  H_{\rm at}&=&\sum_{i=1,2} \left\{\right.
  \omega_r\proj{r_i}+\omega_s\proj{s_i}+\omega_t\proj{t_i}
\nonumber\\
  &&+\omega_1\proj{1_i}
\nonumber\\
  &&+[(\Omega_{ri}/2)\textrm{e}^{-i\omega_{{\rm L}r}t}\op{r_i}{1_i}
  +\textrm{H.c.}]
\nonumber\\
  &&+[(\Omega_{si}/2)\textrm{e}^{-i\omega_{{\rm L}s}t}\op{s_i}{0_i}
  +\textrm{H.c.}]
\nonumber\\
  &&+[(\Omega_{ti}/2)\textrm{e}^{-i\omega_{{\rm L}t}t}\op{t_i}{0_i}
  +\textrm{H.c.}]
  \left.\right\},
\\
  H_{\rm int}&=&\sum_{i=1,2}\left(
  g_r\op{r_i}{0_i}a_i+g_s\op{s_i}{1_i}a_i+\textrm{H.c.}\right),
\end{eqnarray}
(H.c. denotes Hermitian conjugate) and
\begin{eqnarray}
  \mathcal{L}_{\rm cav}\rho_{\rm T}&=&\sum_{i=1,2}\kappa_i
  \damping{a_i}{\rho_{\rm T}}{a_i^\dag}
\nonumber\\
  &&-2\sqrt{\epsilon\kappa_1\kappa_2}
  \brar{\com{a_2^\dag}{a_1\rho_{\rm T}}
  +\com{\rho_{\rm T}a_1^\dag}{a_2}}.
\end{eqnarray}
Here, $a_i$ is the cavity mode annihilation operator for cavity~$i$,
$\omega_{{\rm L}j}$ ($j\in\{ r,s,t\}$) denote the laser frequencies,
and the term $\mathcal{L}_{\rm spon}\rho_{\rm T}$ describes atomic
spontaneous emission.
The term $\mathcal{L}_{\rm cav}\rho_{\rm T}$ describes damping of the
cavity modes through their output mirrors, plus the unidirectional
coupling from cavity 1 to cavity 2 \cite{GardinerCarmichael93}.

Assuming large detunings of the fields from the excited
atomic states
(i.e., $\modulus{\Delta_j}\gg \modulus{\Omega_{ji}},g_{r,s},\kappa_i,
\gamma_j$, where $\gamma_j$ is the linewidth of state $\ket{j}$),
we can adiabatically these states, and neglect atomic spontaneous
emission, to obtain a simplified model of the system in the form of
a reduced master equation for a pair of effective two-level
atoms (states $\ket{0}$ and $\ket{1}$) coupled to the
cavity modes. This reduced system is characterized by the parameters
\begin{eqnarray}
  \beta_{ki}=\frac{g_k\Omega_{ki}}{2\Delta_k} , \;\;\;
  \alpha_{ji} = \frac{|\Omega_{ji}|^2}{4\Delta_j} , \;\;\;
  \eta_k = \frac{g_k^2}{\Delta_k} ,
\end{eqnarray}
where $k\in\{r,s\}$, $i\in\{1,2\}$, and $j\in\{r,s,t\}$;
$\beta_{ki}$ are Raman coupling rates, while $\alpha_{ji}$ and
$\eta_k$ correspond to laser- and cavity-induced atomic level
shifts, respectively.

To further reduce the model, we assume the ``bad-cavity'' limit:
$\kappa_i\gg\modulus{\beta_{ki}},\modulus{\eta_k}$.
This allows us to adiabatically eliminate the cavity mode
to give a master equation for the atomic density matrix $\rho$:
\begin{eqnarray}
\label{eq:me}
  \dot\rho&=&\sum_{i=1,2}\damping{R_i}{\rho}{R_i^\dag}
\nonumber\\*
  &&-2\sqrt{\epsilon}
  \brar{\com{R_1\rho}{R_2^\dag}+\com{R_2}{\rho R_1^\dag}},
\end{eqnarray}
where
$R_i=\brar{\beta_{ri}\op{0_i}{1_i}
+\beta_{si}\op{1_i}{0_i}}/\sqrt{\kappa_i}$.

The first line of (\ref{eq:me}) describes the separate
interaction of each atom with an effective squeezed reservoir
\cite{Clark03},
while the second line describes a unidirectional
coupling between the atoms. As we show below, this combination
of squeezing and coupling facilitates the preparation of an
entangled steady state of the atoms.

Note that the derivation of (\ref{eq:me}) also requires that
the phase of the effective two-level system remains constant
with respect to the phase difference between~$\Omega_{ri}$
and~$\Omega_{si}$, i.e., the two-level atomic systems and
squeezed reservoirs must be ``resonant'' with each other.
Under conditions of Raman resonance
($\omega_{\rm cav}-\omega_{{\rm L}r}=\omega_{{\rm L}s}-
\omega_{\rm cav}=\omega_1$), this requirement leads to the
condition
\begin{equation}
\label{eq:condition}
\alpha_{ri} - \alpha_{si} - \alpha_{ti} = 0 .
\end{equation}
It is to satisfy this condition while retaining flexibility
in our choices of~$\Omega_{ri},\Omega_{si}$ and~$\Delta_{r,s}$
that we use the additional transition $\ket{0}\leftrightarrow\ket{t}$.
The level shift $\alpha_{ti}$ provides an
extra degree of freedom with which to satisfy~(\ref{eq:condition}).

If the atoms are driven such that
$\beta_{r1}/\sqrt{\kappa_1}=\beta_{r2}/\sqrt{\kappa_2}=a$
and
$\beta_{s1}/\sqrt{\kappa_1}=\beta_{s2}/\sqrt{\kappa_2}=b$
then an analytic steady-state solution of
(\ref{eq:me}) can be obtained as
\begin{equation} 
  \label{eq:rhoss}
  \rho_{\rm ss}=\brar{\begin{array}{cccc}
  \rho_{11}&0&0&\rho_{14}\\
  0&\rho_{22}&\rho_{23}&0\\
  0&\rho_{23}^\ast&\rho_{33}&0
  \\\rho_{14}^\ast&0&0&\rho_{44}\\\end{array}},
\end{equation}
in the basis
$\brac{\ket{1_11_2},\ket{1_10_2},\ket{0_11_2},\ket{0_10_2}}$,
where
\begin{eqnarray}
  \rho_{11}&=&
  \brar{\mbs+\brar{1+\ep-4\ep^2}\mat\mbf+\ep\mbt\maf}/D,
\nonumber\\
  \rho_{22}&=&
  \mat\mbt(1-\ep)\brar{\mat+(1+4\ep)\mbt}/D,
\nonumber\\
  \rho_{33}&=&
  \mat\mbt(1-\ep)\brar{\mbt+(1+4\ep)\mat}/D,
\nonumber\\
  \rho_{44}&=&
  \brar{\mas+\ep\mat\mbf+\brar{1+\ep-4\ep^2}\mbt\maf}/D,
\nonumber\\
  \rho_{14}&=&
  \sqrt{\ep}a^\ast b\brar{\maf+(2-4\ep)\mat\mbt+\mbf}/D,
\nonumber\\
  \rho_{23}&=&
  2\sqrt{\ep}(1-\ep)\mat\mbt\left(\mat+\mbt\right)/D,
\end{eqnarray}
and
\begin{eqnarray*}
D=(\maf+\mbf+2(1+2\ep-4\ep^2)\mat\mbt )(\mat+\mbt ).
\end{eqnarray*}

In general, $\rho_{\rm ss}$ describes an entangled mixed state, but
for the case of ideal coupling between cavities ($\epsilon=1$)
we obtain the \emph{pure state} $\rho_{\rm ss}=\proj{\psi}$, where
\begin{equation} \label{eq:psi_ss}
  \ket{\psi}=\frac{a\ket{0_10_2}+b\ket{1_11_2}}
  {\sqrt{\modsquared{a}+\modsquared{b}}}.
\end{equation}
Note that the generation of this
pure state coincides with a complete absence of photons in the
output field from cavity 2, i.e.,
the cascaded system as a whole is prepared in a {\em dark}, or
{\em decoherence-free} state.

The state (\ref{eq:psi_ss}) approximates the maximally-entangled
Bell states
$\ket{\phi^\pm}=\brar{\ket{0_10_2}\pm\ket{1_11_2}}/\sqrt{2}$
in the limit that $a\simeq\pm b$.
The Bell states
$\ket{\psi^\pm}=\brar{\ket{0_11_2}\pm\ket{1_10_2}}/\sqrt{2}$
may be approximated in the steady state in the same limits
simply by choosing
$\beta_{r1}/\sqrt{\kappa_1}=\beta_{s2}/\sqrt{\kappa_2}=a$
and
$\beta_{s1}/\sqrt{\kappa_1}=\beta_{r2}/\sqrt{\kappa_2}=b$.

To gauge the performance of the scheme under more general conditions,
we have performed a variety of numerical simulations taking into
account the dynamics of the cavity mode, imperfect coupling between
the cavities, and the effects of atomic spontaneous emission.
To quantify the degree to which the scheme
generates a maximally entangled state, we use the
measure of fidelity (maximal singlet fraction), for which an analytic
form exists in the case of two qubits \cite{Badziag00}.

\begin{figure}
\begin{center}
\includegraphics[scale=0.4]{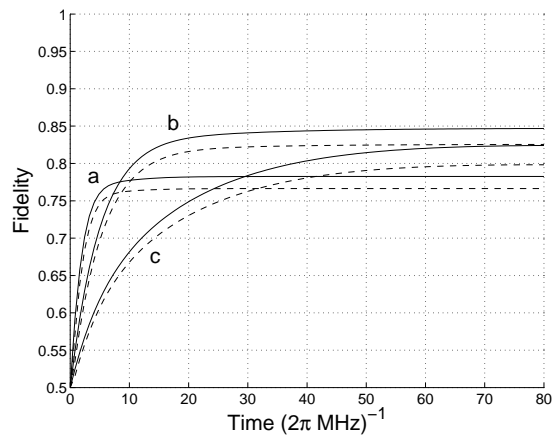}
\caption{Fidelity (maximal overlap with a maximally
entangled state) versus time for
$\brar{g,\kappa,\gamma,\Delta,\Omega_s}/2\pi=
\brar{110,14.2,5.2,8000,100}\textrm{MHz}$,
$\epsilon=0.98$, and
(a) $a/b=3$, (b) $a/b=2$, (c) $a/b=1.5$.
Solid lines: from Eq.~(\ref{eq:me}).
Dashed lines: cavity dynamics and spontaneous emission included.}
\label{fig:fid_time3}
\end{center}
\end{figure}

The evolution of the atoms (each prepared initially in the
state~$\ket{0}$) towards a highly entangled state is shown in
Fig.~\ref{fig:fid_time3} in a plot of fidelity against time,
for several values of the ratio $a/b$ and for a non-unit coupling
efficiency $\epsilon$. The cavity-QED
parameters ($g$, $\kappa$, $\gamma$) used are taken from a recent
experiment \cite{Hood00};
for simplicity, we assume $g_r=g_s=g$.
The solid lines are derived from solutions
to (\ref{eq:me}), while the dashed lines are derived from a more
complete model including both cavity dynamics and the effects of
atomic spontaneous emission from the three excited levels
$\ket{r_i}$, $\ket{s_i}$ and $\ket{t_i}$. (We assume
equal branching ratios where two different decay channels are
possible, e.g., $\ket{r_i}\rightarrow\ket{0_i}$ and
$\ket{r_i}\rightarrow\ket{1_i}$.)
With the cavity dynamics included in the model, one finds that
cavity-induced level shifts (proportional to $\eta_{r,s}$) can
play an appreciable role; in particular, when they are not
substantially less than $\kappa$.
These shifts can be compensated for, and the fidelity of the prepared
state optimized, by choosing $\eta_r=\eta_s\equiv\eta$,
$\omega_{\rm cav}=\frac{1}{2}(\omega_{{\rm L}s}+\omega_{{\rm L}r})
-\eta$, and $\alpha_{ri}-\alpha_{si}-\alpha_{ti}=
\frac{1}{2}(\omega_{{\rm L}s}-\omega_{{\rm L}r})-\omega_1$, as
we do for the results presented in Figs.~3--5.

Returning to Fig.~3, we note first the slowing-down of the
evolution towards the steady state as the ratio $a/b$ approaches
unity. This behavior is characteristic of atomic evolution in a
squeezed reservoir as the degree of squeezing increases
\cite{Clark03}, which here corresponds to
$a/b\rightarrow 1$. In fact, the slowest timescale in the atomic
dynamics scales in proportion to $(a/b-1)^{-2}$, which limits the
maximum attainable fidelity once spontaneous emission is
taken into account. As $a/b\rightarrow 1$ the scheme also becomes
more sensitive to losses in transmission between the cavities
(i.e., $\epsilon <1$). This is highlighted by the fact that the
solid curve for $a/b=1.5$ lies below that for $a/b=2$ (contrary to
the ideal case when $\epsilon =1$).

\begin{figure}
\begin{center}
\includegraphics[scale=0.4]{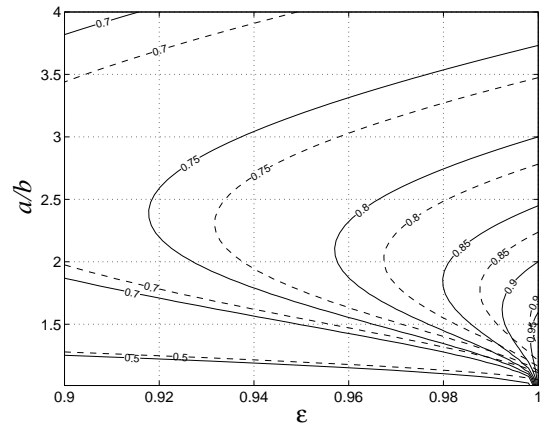}
\caption{Steady state fidelity as a function of $\epsilon$
and the ratio $a/b$. Other than $\Omega_r$, $\Omega_t$, and $\epsilon$,
the parameters are the same as in Fig.~\ref{fig:fid_time3}.
Solid lines: from Eq.~(\ref{eq:me}).
Dashed lines: cavity dynamics and spontaneous emission included.}
\label{fig:contours}
\end{center}
\end{figure}

These features are illustrated further in the contour
plot of Fig.~\ref{fig:contours}, which shows the
steady-state fidelity as a function of $a/b$ and $\epsilon$.
Importantly, this plot also demonstrates that significant
steady state entanglement is possible for relatively modest values
of $a/b$ and $\epsilon$. Note in addition that the characteristic
state preparation times (see Fig.~3) are typically orders of
magnitude smaller than achievable single atom
trapping times (see, e.g., \cite{McKeever03}).

\begin{figure}
\begin{center}
\includegraphics[scale=0.4]{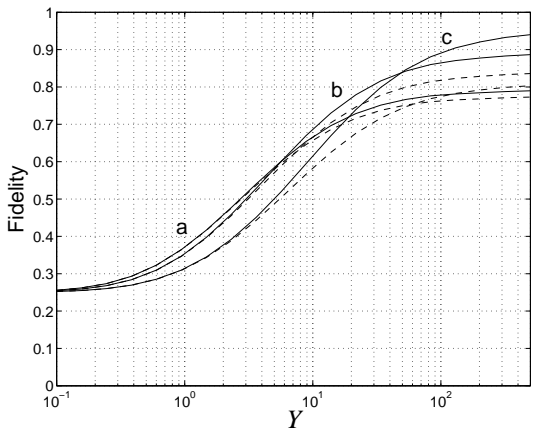}
\caption{Steady state fidelity versus $Y=g^2/(\kappa\gamma )$ for
(a) $a/b=3$, (b) $a/b=2$, and (c) $a/b=1.5$, with
$\epsilon=1$ (solid lines) and $\epsilon=0.98$ (dashed lines).
To obtain these curves, $g$ is varied, while
$\{\beta_{ri},\beta_{si}\}$ are kept constant by adjusting
$\{\Omega_{ri},\Omega_{si}\}$ appropriately (so that the
condition $\kappa\gg\beta_{ri},\beta_{si}$ remains satisfied).}
\label{fig:fid_coop}
\end{center}
\end{figure}

A closer examination of rates associated with Eq.~(\ref{eq:me})
and rates associated with atomic spontaneous emission
($\sim \gamma\Omega_j^2/\Delta_j^2$) shows that
the effects of spontaneous emission can be obviated, for a
particular value of $a/b$, with a
sufficiently large value of $g^2/(\kappa\gamma )$.
To quantify this more carefully,
the steady state fidelity is plotted in Fig.~\ref{fig:fid_coop}
against the cooperativity parameter $Y=g^2/(\kappa\gamma )$
for several values of $a/b$ and for
two values of the coupling efficiency $\epsilon$.
The effects of spontaneous emission are clearly suppressed
for $Y\gg 1$, although this condition becomes more demanding
as $a/b\rightarrow 1$. However, it is also apparent from Fig.~5
(and Figs.~3 and 4) that for a particular $\epsilon <1$ there
exists an optimum value of $a/b$, greater than one, for which the
achievable fidelity is maximized.

By breaking the symmetry between the two atoms with respect to
Raman driving strengths (i.e., by varying the ratios
$\beta_{r1}/\beta_{r2}$, $\beta_{s1}/\beta_{s2}$, as well as
$\beta_{r1}/\beta_{s1}$ and $\beta_{r2}/\beta_{s2}$), it is
possible in principle to generate a wide variety of mixed
entangled states, corresponding to most of the allowed
combinations of entropy and concurrence \cite{Gu03,Clark03}.
Given multiple atoms within each cavity and the ability to
address these atoms individually and sequentially with laser
fields, one might also contemplate the preparation of multiple
pairs of entangled atoms, to which one could apply entanglement
purification procedures \cite{Dur99}.
Alternatively, with multiple atoms coupled collectively to the
cavity mode at each site, it should be possible to prepare
entangled states of separated atomic ensembles
\cite{Duan00,Duan01,Duan02}.
The collective enhancement of the atom-cavity coupling strength
associated with many-atom systems would also alleviate the
need for strong single-atom cavity coupling strength (i.e., the
condition $g^2/(\kappa\gamma)\gg 1$ would become
$Ng^2/(\kappa\gamma)\gg 1$, where $N$ is the number of atoms).

In conclusion, we have proposed a scheme for the
unconditional steady-state preparation of
entangled states of distantly separated atoms. The scheme
does not require entangled light fields or projective
measurements, and appears to be feasible with
existing experimental parameters.

\end{document}